\documentstyle[11pt,newpasp,twoside,epsf]{article}
\begin{document}
\title{Observations of possible symbiotic star V335 Vul}

\author{Petr Sobotka, Ondrej Pejcha}
\affil{MEDUZA, c/o N. Copernicus Observatory and planetarium in Brno, Kravi hora 2, 616 00 Brno, Czech Republic}
\author{Ladislav Smelcer}
\affil{MEDUZA, Valasske Mezirici Observatory, Vsetinska 78, 757 01 Valasske Mezirici, Czech Republic}
\author{Pavel Marek}
\affil{MEDUZA, Skroupova 957, 500 02 Hradec Kralove, Czech Republic}

\begin{abstract}
We present visual and CCD VRI observations of a suspected symbiotic binary V335 Vul. Long term photometry suggests 
usual Mira or SRa pulsations. High speed $V$ band CCD photometry didn't reveal anything unusual above 0.02 mag.
\end{abstract}

\section{Introduction}
V335~Vul = (AS~356 = GSC 02128.00676 = LD~120 = IRAS 19211+2421) was first noticed by Merril and Burwell (1950) as a 
star with an apparent H-$\alpha$ emission. Nassau and Blanco (1957) resolved V335~Vul as a cool carbon star. This was confirmed 
by Stephenson (1971).

The star was found to be variable by Collins (1991) on photographic plates with maximum magnitude at 10.1 mag and minimum
under the limiting magnitude of the plates at 12.7 mag. Collins also determined the type of variability to be SR with 
ephemeris like Max = 2447767 + 336 x $E$. This information caused the appearance of the star in the 71$^{st}$ name-list of 
variable stars (Kazarovets et al. 1993). The independent discovery of variability came from Dahlmark (1993). From his 
photographic measurements between 1985 and 1992, he could derive following ephemeris: 2446740 + 342.0 x $E$. He also noted 
that the period of light changes was stable during 1967 and 1992 and found $B-V$ = 3.3 mag. The type of variability was 
determined to be on the edge between Mira and SRa.

Munari, Tomov and Rejkuba (1999) argued (using echellete spectra) that V335~Vul is probably a symbiotic binary. Munari 
et al. (1999) discovered an apparent outburst of V335~Vul with significant change of colour index $B-V$ from 5.01 to 3.06 
mag. Belczynski et al. (2000) attributed V335~Vul to stars suspected from symbiotic nature. Munari and Jurdana-Sepic 
(2002) examined Asiago photographic plates for data on V335~Vul, but insufficient coverage of two years by only 26 points
in $B$ or $V$ filters did not permit detailed analysis.

\section{Observations}
Since the spring of 2000, MEDUZA group has been conducting extensive observing campaign on V335~Vul. This include 
long-term VRI CCD measurements from N. Copernicus Observatory and planetarium in Brno by P. Sobotka and O. Pejcha, 
long-term $V$ band CCD measurements conducted by L. Smelcer at Valasske Mezirici Observatory, high-speed $V$ band CCD 
photometry on three nights made at Hradec Kralove Observatory by P. Marek, and visual observations carried out by 
amateur astronomers in the Czech and Slovak Republics, mostly by P. A. Dubovsky, L. Brat and O. Pejcha. As a result 
of the campaign, 1086 CCD long-term measurements in all filters and 192 visual estimates have been made.

CCD observations from Brno and Hradec Kralove observatories were reduced by the means of Munidos software (Hroch, 
Novak and Kral, 2002); observations from Valasske Mezirici were reduced using CCDOPS software bundled with SBIG CCD
cameras. CCD data from Brno Observatory have been transformed to the standard Johnson-Cousins system, CCD data from 
Valasske Mezirici Observatory have not. Systematical error of the transformations should not exceed 0.1 mag level. 

All observations were made with respect to the precisely measured comparison stars of Henden and Munari (2000). The only exceptions are the two brightest
stars used only for visual observations - their magnitudes were taken from the Tycho and Tycho-2 catalogues.

Before doing any analysis, visual observations have been averaged by moving averages (with 50 days width and 1 day step)
using MedPrum procedure written by M. Haltuf. For plot clarity, weighted averages of CCD observations made on the 
same night with the same filter and by one observer were done. Times of minima and maxima were derived by Kwee and Van 
Woerden method using program AVE by R. Barbera.

\section{Results}
High-speed photometry did not reveal anything unusual above 0.02 mag level. This is not unexpected - amplitudes of 
flickering in the $V$ band tend to be very small among symbiotic binaries. This is not the best way how to prove or disprove
the symbiotic nature of the star.

\begin{figure}
\plotone{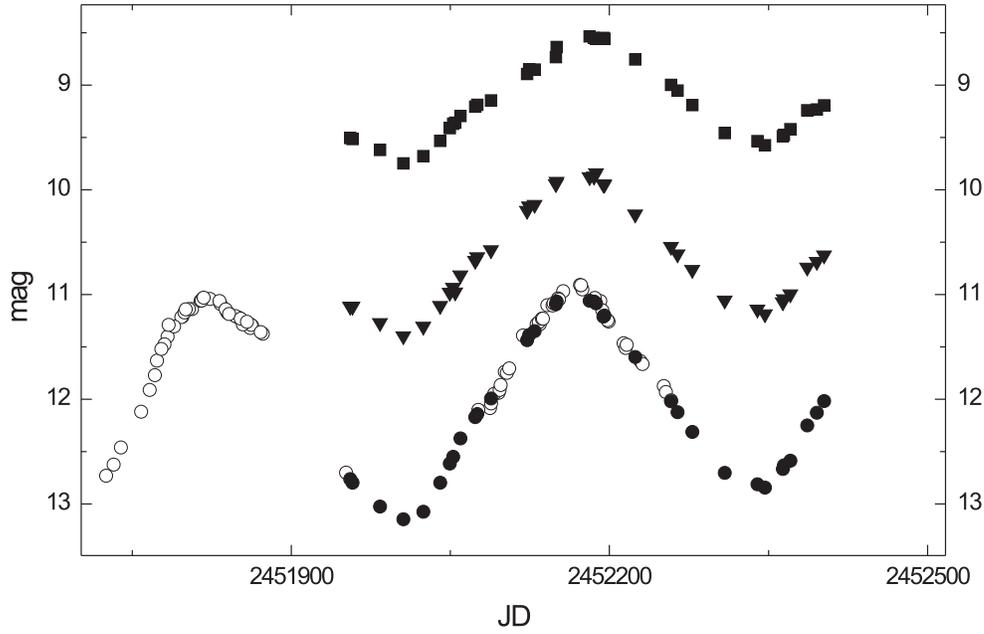}
\caption{CCD observations of V335~Vul. Squares represents CCD+I, triangles represents CCD+R and circles represents CCD+V
data, respectively. Open circles represents data from Valasske Mezirici and filled circles
represents data from Brno.}
\end{figure}

In Figure 1 we present averaged VRI light curves of V335~Vul as obtained from the multi-colour observations by P. Sobotka 
and O. Pejcha (and also L. Smelcer in the case of $V$ filter). All our observations suggest ordinary Mira or SRa type 
variations with slightly different depths of minima. The observed maximum is symmetric in all passbands. Maximum and 
minimum brightness vary in time, see Table 1.

\begin{figure}
\plotone{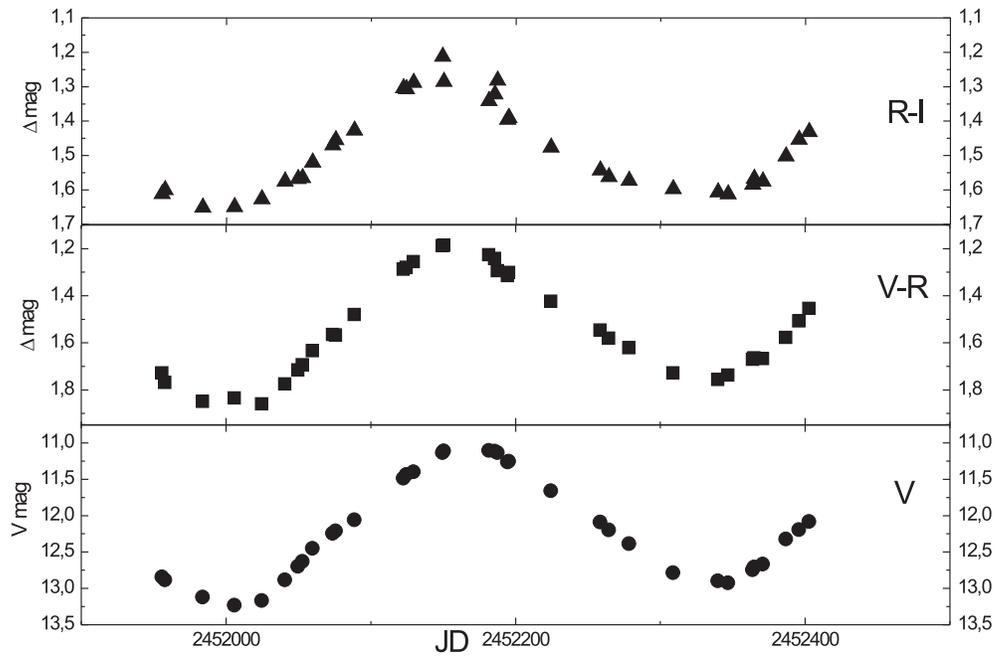}
\caption{CCD observations of V335~Vul. Changes of $R-I$ and $V-R$ colour indices compared with V band light changes are presented.}
\end{figure}

\begin{table}
\caption{Magnitudes of minima and maxima and average amplitudes in filters.}
\begin{center}
\begin{tabular}{lccccc}
\tableline\tableline
\noalign{\smallskip}
filter & max$_1$ & min$_1$ & max$_2$& min$_2$ & $\Delta$ $mag$\\
\noalign{\smallskip}
\tableline
\noalign{\smallskip}
$V$ & 11.23 & 13.23 & 11.04 & 12.92 & 1.94\\
$R$ & & 11.40 & 9.82 & 11.19 & 1.48\\
$I$ & & 9.75 & 8.52 & 9.58 & 1.15\\
\noalign{\smallskip}
\tableline
\tableline
\end{tabular}
\end{center}
\end{table}

We conclude that the outburst announced by Munari et al. (1999) is identical with ordinary Mira-like pulsations. If V335~Vul
is a symbiotic binary, it is similar e.g. to R Aqr.

In Figure 2 we present changes of $R-I$ and $V-R$ colour indices compared with $V$ band light changes. As expected, the star 
is redder near minimum light. Changes of colour indices are presented in Table 2.

\begin{table}
\caption{Magnitudes of minima and maxima of colour indices.}
\begin{center}
\begin{tabular}{lcccc}
\tableline\tableline
\noalign{\smallskip}
index & min$_1$ & max & min$_2$ & $\Delta$ $mag$\\
\noalign{\smallskip}
\tableline
\noalign{\smallskip}
$V-R$ & 1.86 & 1.19 & 1.75 & 0.62\\
$R-I$ & 1.65 & 1.25 & 1.61 & 0.38\\
\noalign{\smallskip}
\tableline
\tableline
\end{tabular}
\end{center}
\end{table}

Times of maximum depends on wavelength: $V = 2452169 \pm 1$, $R = 2452173 \pm 1$ and $I = 2452183 \pm 1$. 
Maxima of $R$ and $I$ band curves occurred 4 and 14 days after the $V$ band maximum, respectively.

Observed $V$ band maxima occurred at JD $2451822.2 \pm 0.5$ and $2452169 \pm 1$ leading to the period of approx. 347 days, which 
is similar to previous works.

All visual observations as well as the chart with sequence can be downloaded via MEDUZA webpage at 
http://www.meduza.info.

\begin{acknowledgments}
We are thankful to all visual and CCD observers who dedicated their time to V335~Vul and L. Kral and M. Haltuf for software
support. This research has made use of NASA's Astrophysics Data System.
\end{acknowledgments}

\end{document}